# La leggenda del quanto centenario

Anna Maria Aloisi

IPSIA "A.Meucci", Cagliari, IT

http:apmf.interfree.it, ampf@interfree.it,

aloisiannamaria@progettomarte.net

Pier Franco Nali

*Servizio per l'Innovazione Tecnologica e per le Tecnologie dell'Informazione e delle Comunicazioni ,*

*Regione Sardegna, Cagliari, IT*

pnali@regione.sardegna.it

Nella storia della fisica del XX secolo la scoperta del quanto d'energia e della discontinuità quantistica ha segnato l'inizio di quello straordinario processo di sviluppo, che a buon diritto può esser definito "rivoluzione quantistica", che ha trasformato profondamente ed in modo permanente la nostra immagine del mondo fisico.

Tradizionalmente, la nascita della teoria quantistica viene collegata alla dimostrazione presentata da Planck alla Società Tedesca di Fisica, durante la seduta del 14 Dicembre 1900, della famosa formula della radiazione termica. Per ottenere la sua fondamentale dimostrazione, Planck aveva dovuto introdurre l'ipotesi che l'energia potesse assumere valori discreti, e questo viene considerato l'inizio della teoria quantistica.

In anni recenti, anche per la concomitanza con le celebrazioni per il centenario della scoperta di Planck (2000), numerosi studiosi[1] hanno ripreso in esame le ricostruzioni storiche di

---

[1] Per citarne soltanto alcuni: P. Cerreta, *La nascita dei quanti: storiografie a confronto*, in «Atti del XVI Congresso Nazionale di Storia della Fisica e dell'Astronomia», Como 1996; J. Büttner, O. Darrigol, D. Hoffmann, J. Renn, M. Schemmel, *Revisiting the Quantum Discontinuity*, Max-Planck-Institut für Wissenschaftsgeschichte, Berlin, August 2000 [http://www.mpiwg-berlin.mpg.de/Preprints/P150.PDF]; H. Kragh, *Max Planck: the reluctant revolutionary*, Physics World, dicembre 2000 [http://physicsweb.org/artiche/world/13/12/8]; G. Parisi, *Planck's Legacy to Statistical Mechanics*, intervento alla Conferenza «Max Planck: l'inizio della nuova fisica», Accademia dei Lincei, 6 dicembre 2000 [http://arXiv.org/cond-mat/0101293]; O. Darrigol, *Max Planck, révolutionnaire malgré lui. Continuités et discontinuités dans "l'acte désespéré" de Max Planck*, Conférence ENSTA, 12 décembre 2001; F. La Teana, *La nascita*

quel famoso avvenimento, arricchendo di nuovi contributi il dibattito che da lungo tempo appassiona e divide gli storici sul vero ruolo di Planck nella nascita della teoria dei quanti. Mentre l'interpretazione convenzionale da sempre attribuisce a Planck il pieno merito di aver posto le basi della teoria quantistica, una tesi alternativa – avanzata oltre venticinque anni fa dallo storico della scienza Thomas Kuhn[2] e riproposta di recente sulla scorta di nuovi elementi – ne ha messo in ombra il lavoro e i risultati, sostenendo che la teoria quantistica non nacque in quel famoso giorno di dicembre del 1900, ma quando Einstein presentò l'ipotesi dei quanti di luce, nel marzo del suo "*annus mirabilis*" 1905, esattamente cento anni fa.

In questo "Anno mondiale della fisica 2005" abbiamo così la ghiotta occasione di celebrare, insieme al centenario della relatività, anche il "vero" centenario dei quanti.

È bene dire subito che ambedue le interpretazioni sono molto controverse. Per entrambe vi sono elementi a favore e contro e il dibattito rimane aperto. Su un punto, però, tutti gli storici concordano: il resoconto degli avvenimenti che viene presentato in molti manuali scolastici e testi divulgativi di fisica[3] è una narrazione mitica, più vicina ad una favola che alla verità storica. Stando a questa versione dei fatti, la teoria quantistica apparve quando ci si rese conto che le previsioni della fisica classica sulla distribuzione energetica della radiazione termica erano in grave disaccordo coi risultati degli esperimenti.

Si tratta di una leggenda. In realtà la teoria quantistica non è nata in seguito a qualche frattura nella fisica classica. La sua origine va ricercata in un lungo processo di sviluppo, avvenuto tra la metà del XIX e gli inizi del XX secolo, legato nelle sue fasi iniziali agli studi sulla radiazione termica connessi col secondo principio della termodinamica, e agli intrecci del problema della radiazione con altri problemi fondamentali di meccanica, elettrodinamica e termodinamica.

### *Più nero del nero*

Gli studi sulla radiazione termica – o radiazione di corpo nero – fecero un grande passo in avanti nel 1859, quando Kirchhoff ne congetturò la natura fondamentale, ipotizzando che non dipendesse dalle proprietà dei corpi radianti e che l'andamento della distribuzione spettrale con la frequenza avesse la forma di una funzione universale della frequenza *f* e della temperatura *T*: $u = u(f,T)$.

Un corpo nero, secondo la definizione che ne diede Kirchhoff stesso, è un corpo che assorbe tutte le radiazioni, e portato all'incandescenza le emette tutte; il suo spettro di emissione ha la forma caratteristica schematizzata nella figura sottostante.

---

*del concetto di quanto*, nei «Quaderni di Fisica Teorica», Bibliopolis, Napoli 2002; Arpita Roy, *Black-Body Problematic: Genesis and Structure*, in «Theory and Science», Vol. 3, Issue 1, Winter 2002; Luis J. Boya, *The Termal Radiation Formula of Planck (1900)*, in «Rev. Real Accademia de Ciencias Zaragoza», 58(2003):91-114 [http://arXiv.org/physics/0402064]; A. Maccari, *Planck e la nascita della fisica dei quanti*, in «Giornale di Fisica», Vol. XLV, N. 1, Gennaio-Marzo 2004.

[2] T. S. Kuhn, *Black-Body Theory and the Quantum Discontinuity: 1894-1912*, Clarendon Press, Oxford 1978 [trad. it. *Alle origini della fisica contemporanea. La teoria del corpo nero e la discontinuità quantica*, Il Mulino, Bologna 1981].

[3] Uno per tutti il famoso libro di George Gamow, *Trent'anni che sconvolsero la fisica. La storia della teoria dei quanti*, Zanichelli, Bologna 1966.

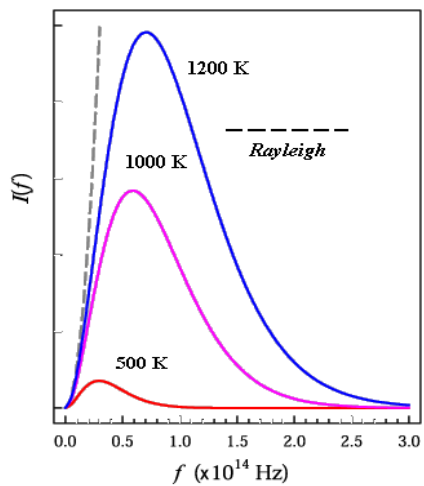

Un corpo nero è perciò quanto di più nero si passa immaginare (assorbe tutte le radiazioni incidenti). Si tratta, naturalmente, di una idealizzazione. Una situazione molto vicina al comportamento di un corpo nero ideale si trova all'interno di una cavità con pareti a temperatura omogenea, che comunica con l'esterno attraverso un piccolissimo foro; la radiazione esterna che cade nel foro finisce dentro la cavità, rimanendovi imprigionata.

La radiazione all'interno di una cavità ideale (radiatore a cavità)[4] risulta avere la stessa costituzione di quella emessa da un corpo nero ideale alla stessa temperatura.

Verso la metà degli anni 90 del XIX secolo a Berlino vennero compiute delle misurazioni molto precise, che permisero di determinare sperimentalmente con una notevole accuratezza la funzione *u*, che rappresenta la densità di energia della radiazione termica per unità di frequenza. Nel 1896 Wien trovò per la densità spettrale *u* una formula,

$$u = \frac{af^3}{e^{\frac{bf}{T}}}$$

(con le costanti *a* e *b* da determinare sperimentalmente), che riproduceva molto bene i risultati sperimentali allora noti.

Nel 1899 Planck, partendo da un'espressione dell'entropia di un oscillatore ideale elementare, riuscì a dedurre teoricamente la formula di Wien (che per questo motivo è conosciuta anche come formula di Wien-Planck). Ma poco tempo dopo nuove misure più accurate – eseguite anche queste a Berlino – mostrarono che la formula di Wien-Planck non descriveva correttamente lo spettro a frequenze molto basse.

Nel giugno 1900 Rayleigh congetturò una dipendenza della funzione *u* da $(f/T)^{-1}$ e ottenne, applicando agli oscillatori del corpo nero il cosiddetto teorema di equipartizione dell'energia, una distribuzione energetica della radiazione proporzionale al quadrato della frequenza.

Il teorema di equipartizione, applicato agli oscillatori che costituiscono un corpo nero, afferma che scomponendo il moto dell'oscillatore in tre moti oscillatori lineari, perpendicolari fra di loro, il valor medio dell'energia del moto oscillatorio lineare è dato da $\overline{U} = kT$ ($k$ è la costante di Boltzmann).

---

[4] v. D. Halliday e R. Resnick, *Fisica (Parte seconda)*, Ambrosiana, Milano 1970, p. 585.

La formula di Rayleigh (che più tardi fu dedotta in modo rigoroso da Jeans e per questo motivo è più conosciuta come la formula di Rayleigh-Jeans[5] $u(f,T) = (8\pi f^2/c^3)kT$ ) spiega molto bene l'andamento osservato alle basse frequenze - dove la formula di Wien non è del tutto corretta - ma alle alte frequenze porta ad una distribuzione totalmente in contrasto con i dati sperimentali.

Il contrasto è così netto che ne derivano conseguenze paradossali: la formula di Rayleigh prevede che la densità spettrale continui a crescere per frequenze sempre più alte, dando luogo ad una densità totale infinita. L'assurdità di tale previsione deriva dall'applicazione alla radiazione del teorema di equipartizione, nato nel contesto della meccanica statistica di Boltzmann per il mondo molecolare. Il paradosso divenne noto col nome di "catastrofe ultravioletta"[6] e venne impropriamente presentato come prova della crisi della fisica classica.

Molti testi di fisica accreditano la tesi della crisi della fisica classica nel passaggio alla teoria quantistica[7], presentando una versione dei fatti che suona più o meno così: volendo a tutti i costi salvare la fisica classica dalla incombente catastrofe, Planck, dopo ripetuti e inutili tentativi, si vide alla fine costretto a ricavare in qualche modo una formula che spiegasse correttamente la distribuzione osservata della radiazione e, nel corso del procedimento matematico, si imbatté suo malgrado - lui che era un "conservatore" - nei quanti. Il racconto di solito si conclude con l'affermazione ad effetto che Planck era un "rivoluzionario riluttante".

La verità storica dice, al contrario, che non ci fu nessuna crisi. La catastrofe ultravioletta non ebbe alcun ruolo nelle prime fasi di sviluppo della teoria quantistica. Come ha giustamente osservato Giorgio Parisi, «il fatto che la meccanica classica porta alla cosiddetta catastrofe ultravioletta fu scoperto solo alcuni anni dopo l'articolo di Planck e non viceversa, come suggerito dal folklore usuale»[8].

La congettura di Rayleigh non esercitò influenza alcuna sul lavoro di Planck[9], che non riteneva fondamentale il teorema di equipartizione dell'energia, poiché derivava dalle idee statistiche di Boltzmann che Planck rifiutava di accettare pienamente[10], e pertanto lo ignorò. Peraltro, i risultati di Rayleigh erano stati presentati in forma incompleta e contenevano degli errori; i risultati corretti furono presentati da Jeans soltanto alcuni anni dopo (1905). Infine, neppure gli stessi Rayleigh e Jeans attribuivano al teorema di equipartizione validità universale.

---

[5] Questa formula è conosciuta anche come legge di Rayleigh-Einstein-Jeans (J.S.W. Rayleigh, Phil. Mag. 49, 539, 1900; id., Nature 72, 54, 1905; id., Nature 72, 243, 1905; J.H. Jeans, Phil. Mag. 10, 91, 1905; A. Einstein, Ann. Phys. 17, 132, 1905).

[6] Il termine "catastrofe ultravioletta" fu coniato da Paul Ehrenfest nel 1911.

[7] Per esempio, nel quarto volume del corso «La Fisica di Berkeley» (*Fisica Quantistica* di Eyvind H. Wichmann, Zanichelli, Bologna 1973) si legge: «Nel 1900, prima della scoperta di Planck, la situazione era in realtà disperata. L'applicazione della meccanica statistica classica aveva condotto all'assurda legge della radiazione del corpo nero, la quale stabilisce che l'intensità irradiata cresce monotonicamente con la frequenza, e quindi l'intensità totale irradiata è infinita; il che significa che la radiazione non può essere in equilibrio con la materia a ogni temperatura».

[8] G. Parisi, cit. nella nota [1].

[9] La formula di Rayleigh è del giugno 1900 ed è citata in un lavoro sperimentale del luglio dello stesso anno di cui Planck era quasi certamente a conoscenza quando presentò la sua proposta in ottobre. Peraltro questi risultati congetturali, ottenuti pochi mesi prima del lavoro di Planck, vennero presentati in forma dettagliata soltanto alcuni anni dopo (1905) e furono perciò ininfluenti sul lavoro di Planck.

[10] A quell'epoca lo scetticismo di Planck non era rivolto tanto verso la meccanica statistica di Boltzmann, dei cui metodi egli stesso fece uso, quanto verso le idee atomistiche che ne erano alla base.

Già nel giugno 1900 Rayleigh scriveva[11] che «per qualche ragione non ancora spiegata la dottrina [dell'equipartizione] non è valida in generale, [ma] sembra possibile che si applichi ai modi più gravi» (ossia alle frequenze più basse), e che «la crescita della densità differenziale d'energia con la frequenza [...] va limitata con l'aggiunta di un fattore d'abbattimento» esponenziale, che portò Rayleigh ad esprimere la legge di radiazione come $u(f,T) = c_1 f^2 T e^{-c_2 f/T}$ (con $c_1$ e $c_2$ costanti non determinate). Fu questa espressione, inizialmente, a divenire nota come legge di Rayleigh, e ne venne ben presto dimostrata sperimentalmente l'inadeguatezza.

Tutto questo dimostra, in primo luogo, che quella che oggi è conosciuta come la "formula di Rayleigh- Jeans" non apparve neppure nel 1900 e, in secondo luogo, che neanche Rayleigh vedeva all'orizzonte una crisi della fisica classica. Tanto meno poteva preoccuparsene Planck!

Quello che fece fu, piuttosto, costruire una nuova espressione, riveduta e corretta, dell'entropia del singolo oscillatore, che gli consentì di presentare alla Società Tedesca di Fisica, nella seduta del 19 ottobre 1900, la formula esatta: $u = \dfrac{af^3}{e^{\frac{bf}{T}} - 1}$, che approssimava alle alte frequenze la formula di Wien e alle basse frequenze (infrarosse) quella di Rayleigh, e che soprattutto era in straordinario accordo con le misure sperimentali.

Ma Planck non era ancora soddisfatto. Si rendeva conto della necessità di dare una solida base teorica alla sua formula (bisognava ricavare l'espressione dell'entropia a partire da principi fondamentali), e già dal novembre 1900 si mise alla ricerca di una derivazione basata – non trovando alternative migliori - sul concetto probabilistico di entropia di Boltzmann[12] e sull'introduzione dell'"elemento di energia" (*Energieelement*), cioè sull'assunzione che «l'energia totale $E$ degli oscillatori del corpo nero è composta da un numero completamente determinato di parti finite ed uguali [di energia $\varepsilon$]»[13].

Per tutti gli oscillatori che hanno una certa frequenza di oscillazione $f$, l'elemento di energia, o quanto, si ottiene moltiplicando la costante naturale «$h = 6{,}55 \cdot 10^{-27}$ erg·s per la comune frequenza di oscillazione $f$ degli oscillatori». È questa la famosa legge di quantizzazione di Planck: $\varepsilon = hf$. Dividendo poi l'energia totale $E$ per il quanto $\varepsilon$ si trova il numero $P$ di quanti da distribuire sugli $N$ oscillatori che compongono il corpo nero. La costante naturale $h$ (costante di Planck) è legata alla costante sperimentale $b$ della formula della radiazione attraverso la costante di Boltzmann $k$ ($k = 1{,}346 \cdot 10^{-16}$ erg/K): $h = bk$.

---

[11] J.S.W. Rayleigh, Phil. Mag. 49, 539, 1900 (cit. in G. Tagliaferri, *Storia della fisica quantistica. Dalle origini alla meccanica ondulatoria*, Franco Angeli, Milano 1985, p. 31).

[12] Per inciso, fu Planck a stabilire quella che è nota come equazione di Boltzmann: $S = k \log W$, che mette in relazione l'entropia $S$ col "disordine elementare" $W$. Anche il metodo di calcolo di Planck riprende tecniche impiegate da Boltzmann nel 1872 nell'ambito della teoria cinetica dei gas. Sui "quanti di Boltzmann" (e su altri antecedenti dei quanti ancor più antichi) v. E. Bellone, *Caos e armonia. Storia della fisica*, UTET, 2004.

[13] M. Planck, *Ueber das Gesetz der Energieverteilung im Normalspectrum* (*Sulla legge di distribuzione dell'energia in uno spettro normale*), in «Annalen der Physik», Vol. 4, 1901, pp. 553-563.

Come Planck stesso scrisse molti anni dopo (1931), dovette introdurre la strana ipotesi dei quanti in un "atto di disperazione" (*Akt der Verzweiflung*), senza dargli altro credito se non perché conduceva ad un risultato positivo, che egli voleva raggiungere a qualsiasi costo.

Anche se Heisenberg[14], riferisce che Planck confidò al figlio la sua convinzione (a proposito della legge $\varepsilon = hf$ ) di aver fatto una scoperta paragonabile alla teoria di Newton e di aver "visto la luce" nel 1900, nessuno sembrò accorgersi che era accaduta una rivoluzione. Come ha osservato Armin Hermann, «durante i cinque anni successivi [al 1900] non fu pubblicato niente sulla teoria quantistica» [15] e - come ha scritto lo storico danese Helge Kragh - «Planck non fece eccezione, e l'importanza attribuita al suo lavoro è in gran parte frutto di una ricostruzione storica. Mentre la legge della radiazione di Planck fu accettata rapidamente, quella che oggi consideriamo la sua novità concettuale – il suo fondamento nella quantizzazione dell'energia – passò quasi inosservata. Pochissimi fisici manifestarono interesse per la dimostrazione della formula di Planck, e nei primissimi anni del XX secolo nessuno considerò i suoi risultati in contrasto coi fondamenti della fisica classica. Quanto a Planck stesso, fece ogni sforzo per mantenere la sua teoria sul solido terreno della fisica classica che così tanto amava […]. Quanto alla discontinuità quantistica – la proprietà cruciale che l'energia non varia con continuità ma a "salti" – credette per lungo tempo che fosse una sorta di ipotesi matematica, un artefatto che non si riferiva ai reali scambi di energia tra materia e radiazione. Dal suo punto di vista, non c'era ragione di sospettare una frattura nelle leggi della meccanica classica e dell'elettrodinamica» [16].

### *L'uomo che vide la luce*

Fu l'opera di Einstein sulla teoria dei quanti di luce (o fotoni) del marzo 1905 a segnare una svolta, perché aprì una nuova via di ricerca che venne percorsa da numerosissimi fisici, malgrado l'iniziale scetticismo di molti (fra cui lo stesso Planck). Questa nuova strada portò dapprima, attraverso un lento processo di sviluppo culminato con la teoria atomica di Bohr (1913), alla "vecchia" teoria quantistica, e successivamente, con una ulteriore evoluzione negli anni '20 del XX secolo, alla "nuova" meccanica quantistica.

L'intervento del giovane Einstein – aveva allora, nel 1905, appena ventisei anni - si rivelò decisivo. Egli fu il primo a riconoscere la distanza che separava le nuove acquisizioni dalla fisica allora accettata. Nelle prime pagine del suo articolo sui quanti di luce [17], in assoluta antitesi con la teoria ondulatoria che sembrava inattaccabile, avanzava l'ipotesi [18] che «quando un raggio di luce si espande partendo da un punto, l'energia non si distribuisce su volumi sempre più grandi, bensì rimane costituita da un numero finito di quanti di energia localizzati nello spazio e che si muovono senza suddividersi, e che non possono essere emessi o assorbiti parzialmente».

---

[14] W. Heisenberg, *Fisica e filosofia*, Il Saggiatore, Milano 1967.

[15] A. Hermann, *Quadro storico della teoria quantistica* (saggio introduttivo a A. Einstein, *Die Hypothese der Lichtquanten*, Ernst Battenberg Verlag, Stuttgart 1965 [trad. it. *La teoria dei quanti di luce*, Newton Compton Editori, Milano 1975]).

[16] H. Kragh, cit. nella nota [1].

[17] A. Einstein, *Ueber einen die Erzeugung und Verwandlung des Lichtes betreffenden heuristischen Gesichtspunkt* (*Emissione e trasformazione della luce da un punto di vista euristico*), in «Annalen der Physik», Vol. 17, 1905, pp. 132-148. Gli articoli di Einstein si possono trovare in versione italiana in *L'anno memorabile di Einstein. I cinque scritti che hanno rivoluzionato la fisica del Novecento* (a cura di J. Stachel, E. Ioli), E. Dedalo 2001.

[18] Incidentalmente, questa ipotesi ha una relazione con un'altra ipotesi einsteiniana fondamentale, quella della costanza della velocità della luce per tutti gli osservatori inerziali, posta a fondamento della teoria della relatività ristretta. La costanza della velocità della luce è una conseguenza matematica delle equazioni di Maxwell, ma «la teoria di Maxwell sulla luce (e, in generale, qualunque teoria ondulatoria) afferma che l'energia di un raggio luminoso, emesso da una sorgente luminosa, si distribuisce in modo continuo su di un volume sempre crescente» (A. Einstein, cit. nella nota [17]) e perciò è in contrasto con l'ipotesi dei quanti di luce. Per questo motivo Einstein preferì introdurre la costanza della velocità della luce come ipotesi, perché fosse indipendente dalla teoria di Maxwell. Questa può essere considerata la prima connessione tra teoria quantistica e relatività.

L'articolo proseguiva con alcune considerazione su una difficoltà della teoria della radiazione di corpo nero lasciata irrisolta da Planck. Vediamo in che cosa consisteva questa difficoltà.

Einstein ammetteva la concordanza della formula di Planck con tutte le misure - come era generale convinzione - ma non ne condivideva la dimostrazione. Secondo Einstein, se si voleva rimanere dentro i confini della fisica allora accettata, si doveva partire dall'equazione, stabilita in precedenza da Planck in un quadro totalmente classico, che esprime il legame tra la radiazione libera e gli oscillatori che costituiscono la cavità: $u(f,T) = \frac{8\pi f^2}{c^3} U(f,T)$, dove $U$ è l'energia di equilibrio dell'oscillatore di frequenza $f$. Si giungeva allora in modo naturale - sempre lavorando entro il quadro classico ed applicando agli oscillatori il teorema di equipartizione $\overline{U} = U(f,T) = kT$ - alla formula di Rayleigh $u(f,T) = \frac{8\pi f^2}{c^3} kT$ e non a quella di Planck. Viene spontaneo chiedersi perché Planck non avesse scoperto lui stesso questo fatto prima di Einstein, visto che disponeva di tutti gli ingredienti necessari già prima del 1900. La risposta si trova probabilmente nel suo atteggiamento di cui abbiamo parlato in precedenza nei confronti delle idee di Boltzmann.

Per Einstein, dunque, la formula di Planck era in accordo con gli esperimenti ma non era accettabile teoricamente; viceversa, la legge di Rayleigh andava d'accordo con la teoria ma non con gli esperimenti. Perciò, da qui in avanti decise di considerare il problema «in base all'esperienza, senza stabilire nessuna ipotesi teorica nei confronti dell'emissione e della propagazione della radiazione».

Scelse perciò di procedere da un punto di vista fenomenologico ed usò a tale scopo la legge di Wien, che era ben verificata sperimentalmente per valori elevati di $f/T$, cioè per onde di alta frequenza e bassa densità. Dalla legge di Wien ricavò poi un'espressione dell'entropia della radiazione che mostrava come «l'entropia di una radiazione monocromatica di densità abbastanza ridotta varia in funzione del volume, seguendo la stessa legge che vale per l'entropia di un gas ideale»; e più avanti, dopo aver analizzato la relazione entropia-volume, portò a conclusione il ragionamento: «Una radiazione monocromatica di densità ridotta (nei limiti di validità della legge di Wien) si comporta, nell'ambito della termodinamica, come se fosse composta di quanti di energia di grandezza $hf$, indipendenti tra di loro». Ma il passo veramente rivoluzionario sarebbe venuto subito dopo, con la famosa ipotesi euristica: «Se una radiazione monocromatica (di densità sufficientemente ridotta) si comporta, rispetto alla relazione entropia-volume, come un mezzo discontinuo, costituito da quanti di energia della grandezza $hf$, dovremo esaminare l'ipotesi che le leggi di emissione e di trasformazione della luce siano costituite anche loro, come se la luce fosse formata da simili quanti di energia». Ci troviamo di fronte, evidentemente, a un'estensione dell'ipotesi dei quanti dalla radiazione libera ai processi di emissione e assorbimento che coinvolgono la materia.

L'applicazione di questo principio permise ad Einstein di formulare la legge dell'effetto fotoelettrico – che gli varrà il Nobel 1921 – e di gettar luce su vari altri fenomeni di interazione radiazione-materia, che all'epoca erano un rompicapo per i fisici.

Un aspetto forse meno noto, ma molto interessante, di questo primo lavoro di Einstein sulla teoria quantistica è che in esso - come ha fatto notare Max Born - si trova implicitamente contenuto il principio di dualità onda-particella.

Per rendersene conto si consideri il fatto che Einstein, dopo aver ricavato la formula di Rayleigh dalla teoria di Maxwell (ondulatoria) ed aver mostrato che questa formula porta a risultati assurdi alle alte frequenze, la ricava immediatamente dopo della formula di Planck - che era sperimentalmente corretta - come caso limite per valori elevati di $T/f$, cioè per onde di bassa frequenza e alta densità di radiazione, concludendone perciò che «più sono grandi la densità di energia e la lunghezza d'onda [ossia più è piccola la frequenza] più i nostri principi teorici [ossia la teoria ondulatoria di Maxwell] risultano utilizzabili; per quel che riguarda piccole lunghezze d'onda e radiazioni poco dense i medesimi principi sono del tutto inutilizzabili».

In sintesi, nel regime delle alte densità e basse frequenze il comportamento della luce viene correttamente descritto dalla teoria ondulatoria, che risulta invece inapplicabile al di fuori di tale regime.

Più avanti, dopo aver ricavato un'espressione dell'entropia della radiazione dalla legge di Wien – valida per valori elevati di $f/T$, cioè per onde di alta frequenza e bassa densità - fa la considerazione che «l'entropia di una radiazione monocromatica di densità abbastanza ridotta varia in funzione del volume, seguendo la stessa legge che vale per l'entropia di un gas ideale».

Detto in altre parole, in condizioni di bassa densità e alta frequenza la luce si comporta come un gas, cioè mostra una natura corpuscolare.

L'analogia fra la luce a bassa densità e il gas perfetto si spiega considerando che in un gas rarefatto l'interazione tra molecole è trascurabile. Allo stesso modo, in un "gas di fotoni" disperso in un grande volume sono trascurabili gli effetti di interferenza che caratterizzano il comportamento ondulatorio della luce. Inoltre, come sappiamo oggi dalla meccanica quantistica, alle alte frequenze (piccole lunghezze d'onda) i fotoni sono molto localizzati, riducendo ulteriormente le probabilità di interazione.

In conclusione, nel regime di Rayleigh (bassa frequenza e alta densità) la luce è come se fosse formata da onde che si trasformano in particelle passando nel regime di Wien (alta frequenza e bassa densità).

Einstein tornò sui fotoni un anno dopo (marzo 1906) con un articolo[19] che, secondo Kuhn, più ancora di quello del 1905 annuncia la nascita della teoria quantistica[20]. In quell'articolo Einstein dichiarava che nuovi elementi lo avevano indotto a ritenere che «la teoria di Planck fa uso implicito dell'ipotesi dei fotoni». Einstein in questo modo riconosceva piena validità al lavoro di Planck, e in un certo modo gli rendeva l'onore delle armi.

---

[19] A. Einstein, *Theorie der Lichterzeugung und Lichtabsorption* (*Teoria dell'emissione e dell'assorbimento della luce*), in «Annalen der Physik», Vol. 20, 1905, 199-206.

[20] T. S. Kuhn cit. nella nota [2]. Kuhn scrive: «A metà dell'articolo Einstein scriveva: "Dobbiamo, perciò, riconoscere la seguente posizione come fondamentale per la teoria della radiazione di Planck. L'energia di un risonatore elementare può soltanto assumere valori che sono multipli interi di *hf*. Durante l'assorbimento e l'emissione l'energia di un risonatore varia in modo discontinuo per un multiplo intero di *hf*.". Questo brano costituisce la prima affermazione pubblica del fatto che la derivazione di Planck impone una limitazione nel continuum classico degli stati del risonatore. In un certo senso, esso annuncia la nascita della teoria quantistica».

## *Due giganti a confronto*

Quando nacque, dunque, la teoria quantistica, nel dicembre 1900 o nel marzo 1905? Oppure, come sostiene Khun, nel marzo 1906? E chi ne fu l'artefice, Planck o Einstein?

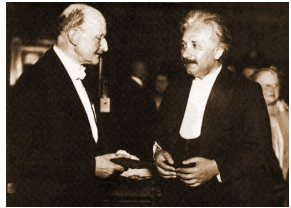

Si potrebbe forse dar credito alla tradizione, e rispondere che fu Planck nel 1900. Si potrebbe allora obiettare che Planck introdusse l'ipotesi dei quanti in un unico passaggio di un lungo ragionamento e poi non ne fece più uso; che la considerò come una sorta di artefatto matematico; che non credeva avesse a che fare con reali scambi radiazione-materia. Ma, dopotutto, Planck ricevette il Nobel nel 1918 proprio per "la sua scoperta dei quanti di energia".

O forse si potrebbe, con maggior convinzione, sostenere che il fondatore della teoria quantistica sia stato Einstein. «I notevoli contributi di Einstein alle prime fasi della teoria quantistica – argomenta Kragh - sono ben noti e indiscutibili. È molto famosa la sua teoria del 1905 dei quanti di luce (o fotoni), ma diede anche importanti contributi nel 1907 sulla teoria quantistica dei calori specifici dei solidi e nel 1909 sulle fluttuazioni di energia[21]. Non vi è dubbio che il giovane Einstein vide più in profondità di Planck, e che soltanto Einstein riconobbe che la discontinuità quantistica era una parte essenziale della teoria di Planck della radiazione di corpo nero. […] La teoria di Einstein dei calori specifici del 1907 fu un elemento importante nel processo che stabilì la teoria quantistica come un campo fondamentale della fisica».

Probabilmente la risposta migliore è che ciascuno di questi due uomini straordinari contribuì in modo fondamentale, sebbene in misura diversa, allo sviluppo della teoria quantistica. Eppure, sorprendentemente, entrambi presero le distanze da questa teoria che tanto avevano contribuito a creare: Planck divenne un paladino della relatività (!), e rimase sempre legato all'amata fisica classica; Einstein considerò sempre la teoria quantistica (e la meccanica quantistica che ne sarebbe scaturita) una teoria provvisoria.

Dalla vicenda che abbiamo narrato possiamo trarre un insegnamento: la scoperta scientifica è un processo complesso, a volte ambiguo e contraddittorio, e non si esaurisce in un unico avvenimento, come quello accaduto in un particolare giorno del dicembre 1900. Attribuire una scoperta a un solo uomo – sia pure un gigante della scienza – è troppo semplicistico. La scoperta è un processo che si protrae nel tempo, non l'intuizione momentanea di un genio.

---

[21] Sugli sviluppi fino al 1917 v. A. Maccari, *Il ruolo di Einstein nello sviluppo della fisica dei quanti*, in «Giornale di Fisica», Vol. XLVI, N. 1, Gennaio-Marzo 2005.